\documentclass[twocolumn,aps,prl,amsmath,amssymb,showpacs,preprintnumbers]{revtex4}

\newcommand{\nc}{\newcommand}

\nc{\be}{\begin{equation}}

\nc{\ee}{\end{equation}}

\nc{\bea}{\begin{eqnarray}}

\nc{\eea}{\end{eqnarray}}

\nc{\xx}{\nonumber\\}

\nc{\ct}{\cite}

\nc{\la}{\label}

\nc{\eq}[1]{(\ref{#1})}

\def\ajou#1&#2(#3){\ \sl#1\bf#2\rm(19#3)}

\def\half{{\frac{1}{2}}}

\def\[{\left [}
\def\]{\right ]}

\begin{document}
\preprint{HU-EP-07/27}
\preprint{KIAS-P07069}

\title{Emergent Gravity And The Cosmological Constant Problem}
\author{Hyun Seok Yang}
\affiliation{School of Physics, Korea Institute for Advanced Study, Seoul 130-012,
Korea \\ Institut f\"ur Physik, Humboldt Universit\"at zu Berlin,
Newtonstra\ss e 15, D-12489 Berlin, Germany}

\date{\today}

\begin{abstract}
We address issues on the origin of gravity and the cosmological
constant problem based on a recent understanding about the
correspondence between noncommutative field theory and gravity. We
suggest that the cosmological constant problem can be resolved in a
natural way if gravity emerges from a gauge theory in noncommutative
spacetime. Especially, we elucidate why the emergent gravity implies
that vacuum energy does not gravitate but only fluctuations around
the vacuum generate gravity. That is, a flat spacetime emerges from
uniform condensation of energy, previously identified with the
cosmological constant.

\end{abstract}


\maketitle

Over the past ten or twenty years, several magnificent astronomical
observations have revealed that our Universe curiously hides its
critical aspects behind the dark side such as dark matter and dark
energy. The experimental data have shown that our Universe is
composed of 4 \% radiations and ordinary matters while 21 \% dark
matter and 75 \% dark energy \ct{cc-review}. The scientific framework of the
twentieth century using the Standard Model and general relativity,
strictly speaking, has failed to shed some light on the dark side in
spite of spectacular success to explain most of phenomena on Earth
and in heaven. Morally speaking, the dark side is thus `the saddest
chapter in theoretical physics' using Heisenberg's word.

The remarkable success of the contemporary theoretical physics has
been based on the action
\begin{equation} \label{einstein-standard}
S= S_G + S_{SM}
\end{equation}
where
\begin{eqnarray} \label{einstein}
&& S_G = - \frac{1}{16 \pi G} \int d^4 x \sqrt{-g} (R + 2 \Lambda), \\
\label{standard}
&& S_{SM} = \int d^4 x \sqrt{-g} {\cal L}_{SM}
\end{eqnarray}
with ${\cal L}_{SM}$ the Standard Model Lagrangian. The equations of
motions for gravitation are given by
\begin{equation} \label{einstein-eq}
R_{\mu\nu}- \frac{1}{2} g_{\mu\nu} R - \Lambda g_{\mu\nu} = - 8 \pi G
T_{\mu\nu},
\end{equation}
where $\Lambda$ is the so-called cosmological constant (CC).
Eq.(\ref{einstein-eq}) contains two fundamental length scales
related to the Planck scale $L_P^2 \equiv 8 \pi G$ and
the length scale $L_H^2 \equiv 1/\Lambda$ defined by the CC and
it shows that the CC acts like a fluid with an energy density given by
\begin{equation} \label{cc}
\rho_\Lambda = \frac{\Lambda}{8 \pi G} = \frac{1}{L_P^2 L_H^2}
\end{equation}
satisfying an exotic equation of state $p= - \rho$. Thus the CC
exerts a negative pressure, causing an exponential expansion of
universe as in inflation era and the current accelerating expansion of
universe.

The CC consists of a uniform and unclustered energy with a negative pressure,
sharing the same property with the dark energy, the most part (75 \%) of
the energy content of our Universe. The CC might be the simplest and
the most natural candidate for the dark energy \ct{cc-review}. So, from the
following, we will identify the CC with the dark energy.

One has realized that the CC can be interpreted as a measure of the
energy density of the vacuum because anything that contributes to
the energy density of the vacuum acts just like a cosmological
constant. One finds that the resulting energy density is of the form
\begin{equation} \label{vacuum-energy}
\rho_\Lambda^{th} = \frac{1}{V} \sum_{\mathbf{k}}
\frac{1}{2} \hbar \omega_{\mathbf{k}} \sim \hbar k_{max}^4
\end{equation}
where $k_{max}$ is a certain momentum cutoff below which an
underlying theory can be trusted.
So we would naturally estimate a contribution to the vacuum energy (\ref{vacuum-energy}),
for example, of order $\rho_\Lambda^{Pl} \sim (10^{18} GeV)^4$
for a quantum field theory of which we can trust all the way up to
the Planck scale $M_{Pl} = (8\pi G)^{-1/2} \sim 10^{18} GeV$.

The observed value of the CC or the dark energy turned out to be very very tiny \ct{cc-review}, say,
\begin{equation} \label{cc-obs}
\rho_\Lambda^{obs} \leq (10^{-12} GeV)^4,
\end{equation}
which is desperately different from the theoretical estimation (\ref{vacuum-energy}).
Their ratio is roughly $\rho_\Lambda^{obs}/\rho_\Lambda^{th} \sim 10^{-120}$, that
would stand for an unprecedent failure in the history of science.
This huge discrepancy between the theoretical and observational
value is the long standing CC problem.

The tiny value of the CC in (\ref{cc-obs}) implies that our Universe
prefers a {\it flat} spacetime. In this sense, the CC problem is how
to understand the dynamical origin of the flat spacetime. But there
is a blind point about the dynamical origin of spacetime in general
relativity; it says nothing about the dynamical origin of flat
spacetime since the flat spacetime is a geometry of special
relativity rather than general relativity. In other words, the flat
spacetime defining a local inertial frame is assumed to be {\it a
priori} given without reference to its dynamical origin. Thus we
raise the first question about whether the CC problem can be solved
within the framework of Einstein gravity. We suspect that it may not
be the case for the following reasons.

The first phenomenological reason is that all attempts to solve the CC problem have been
failed so far (see the reviews \ct{cc-review}). Now there exists some consensus that we need a more
fundamental theory beyond Einstein gravity to resolve the problem
from which gravity is emergent.

The second reason is that gravity and matters respond differently to the vacuum energy.
The equations of motion for matters in Eq.(\ref{einstein-standard}) are invariant
under shifting the matter Lagrangian by a constant $\lambda$:
\begin{equation} \label{shift-symm}
{\cal L}_{SM} \to {\cal L}_{SM} - 2 \lambda.
\end{equation}
However the shift (\ref{shift-symm}) results in that of the energy-momentum tensor of matter
by $T_{\mu\nu} \to T_{\mu\nu} - \lambda g_{\mu\nu}$ in the presence of gravity.
Therefore gravity breaks the shift symmetry (\ref{shift-symm}). The worse is that
such shift (\ref{shift-symm}) allowed by the matter sector changes the CC
by $\Lambda \to \Lambda + \lambda$. Since the vacuum energy (\ref{vacuum-energy}) is originated
from the matter sector, it is very difficult to imagine a definite
solution to the CC problem without cure for this mismatch \ct{pady1}.

The third practical reason is that the near zero CC seems to indicate that the vacuum energy does
{\it not} gravitate. Of course, this conclusion immediately leads to the contradiction with
the principle of general covariance which requires that gravity couples universally
to all kinds of energy. Where is then an exit for the problem ?

In order to find an exit, we may boldly ask a question whether
gravity is really a fundamental force or not. Interestingly, recent
developments in theoretical physics imply a surprising picture about
gravity \ct{review,est-seiberg}:
{\it Gravity is not a fundamental force but a collective phenomenon emergent
from gauge fields}. Let us summarize a few evidences indicating
this remarkable picture.

It has been well-known that there exist thermodynamic descriptions
of gravity \ct{gr-thermo}, which strongly suggests that there are ``atoms of
spacetime" (a microscopic structure of spacetime) as Boltzmann
taught us. As an analogy, the existence of thermodynamics in solids
implies that of discrete constituents, i.e., atoms, and the
elasticity and the density of solid lose their meanings at atomic
level.

Another splendid evidence is coming from the AdS/CFT duality \ct{ads-cft} where
the bulk gravity in higher dimensions emerges from a lower
dimensional large $N$ gauge theory. The AdS/CFT duality is a
thoroughly tested example of the holographic principle \ct{holography} which states
that physical degrees of freedom in gravity resides on a lower
dimensional screen where gauge fields live than gravitational
theories are defined.
Incidentally, the naive estimate \eq{vacuum-energy} in general violates
holographic bounds constrained by the holographic principle \ct{hol-de}.

Finally analogue gravity from condensed matter physics also provides a clear picture on
how (effective) gravity emerges from collective excitations around a Fermi surface and why
the emergent gravity avoids the CC problem \ct{volovik}.
An emerging picture for the CC problem is that
the system in equilibrium adjusts itself such that the energy of
vacuum is zero where
a shift symmetry like Eq.\eq{shift-symm} plays a key role.

As we reasoned above, the CC problem requires a new, radically
different, approach to gravity which essentially has to do with our
understanding of the nature of gravity. We will now examine the CC
problem from the viewpoint of emergent gravity based on the
correspondence between noncommutative (NC) field theory
and gravity \ct{hsy5,hsy7}.

A NC spacetime is obtained by introducing a symplectic
structure $\omega = \half B_{ab} dy^a \wedge dy^b$ and then by quantizing
the spacetime $M$ with its Poisson structure $\theta^{ab} \equiv (B^{-1})^{ab}$,
treating it as a quantum phase space. For example, one can define NC ${\bf R}^{2n}$
by the following commutation relation \ct{sw}
\be \la{nc-spacetime}
[y^a, y^b]_\star = i \theta^{ab}.
\ee

The fact that the NC spacetime \eq{nc-spacetime} is actually
a (NC) phase space  leads to two important consequences \ct{hsy6}:

(I) If we consider a NC ${\bf R}^2$ for simplicity,
any field $\widehat{\phi} \in {\cal A}_\theta $ on the
NC plane can be expanded in terms of the complete operator basis
\begin{equation}\label{matrix-basis}
{\cal A}_\theta = \{ |m \rangle \langle n|, \; n,m = 0,1, \cdots \},
\end{equation}
that is,
\begin{equation}\label{op-matrix}
    \widehat{\phi}(x,y) = \sum_{n,m} M_{mn} |m \rangle \langle n|.
\end{equation}
One can regard $M_{mn}$ in \eq{op-matrix} as components of an $N \times N$
matrix $M$ in the $N \to \infty$ limit.
We then get the relation:
\be \la{sun-sdiff}
\mathrm{Any \; field \; on \; NC} \; {\bf R}^2 \; \cong \; N \times N \;
\mathrm{matrix} \; \mathrm{at} \; N \to \infty.
\ee
If $\widehat{\phi}$ is a real field, then $M$ should be a Hermitian matrix.
The relation \eq{sun-sdiff} means that NC fields can be regarded
as master fields of large $N$ matrices \ct{master}.

(II) An important fact is that translations in NC directions are an inner automorphism of
NC C*-algebra ${\cal A}_\theta$, i.e.,
$e^{ik \cdot y} \star f(y) \star e^{-ik \cdot y} = f(y +  \theta \cdot k)$
for any $f(y) \in {\cal A}_\theta$ or, in its infinitesimal form,
\be \la{inner-der}
[y^a, f]_\star = i \theta^{ab} \partial_b f.
\ee
In the presence of gauge fields, the coordinates $y^a$
should be promoted to the covariant coordinates defined by
\begin{equation}\label{cov-coord}
 x^a (y) \equiv y^a + \theta^{ab} \widehat{A}_b (y)
\end{equation}
in order for star multiplications to preserve the gauge covariance \ct{madore}.
The inner derivations \eq{inner-der} are accordingly covariantized
too as follows
\bea \la{vector-map}
{\rm ad}_{x^a}[f] &\equiv& [x^a, f(y)]_\star = i\theta^{\alpha\beta}
\frac{\partial x^a}{\partial y^\alpha}\frac{\partial f}{\partial y^\beta} + \cdots \xx
&\equiv& V_a^\alpha(y) \partial_\alpha f(y) + {\cal O}(\theta^3).
\eea
It turns out \ct{hsy3,hsy4} that the vector fields $V_a (y) \equiv V_a^\alpha (y) \partial_\alpha$
form an orthonormal frame and hence define vielbeins of a
gravitational metric.

Of course, the pictures (I) and (II) should refer to the same physics, which is essentially an
equivalent statement with the large $N$ duality in string theory.
The $1/N$ expansion in the picture (I) corresponds to the derivative corrections
in terms of $\theta$ for the picture (II).

The above pictures (I) and (II) imply that a field theory on NC
spacetime should be regarded as a theory of gravity, which we refer
to as the emergent gravity \ct{hsy4}. The correspondence between NC field
theory and gravity can be explicitly confirmed for the self-dual sectors
of NC gauge theories.
Recently it was shown in \ct{hsy3}  that self-dual electromagnetism in
NC spacetime is equivalent to self-dual Einstein gravity.
For example, $U(1)$ instantons in NC spacetime are actually gravitational
instantons \ct{hsy12}. The emergent gravity for general cases
can be clarified by systematically applying to the NC gauge theories
the pictures (I) and (II). Let us briefly summarize the construction in \ct{hsy6}.

Consider a NC $U(1)$ gauge theory on ${\bf R}^D = {\bf R}^d_C \times
{\bf R}^{2n}_{NC}$ whose NC part satisfies the relation \eq{nc-spacetime}.
Let us decompose $D$-dimensional $U(1)$ gauge fields as follows:
$A_M (z,y) =(A_\mu, \Phi^a) (z,y) \; (M=1,\cdots, D; \;
\mu=1, \cdots, d; \; a = 1, \cdots, 2n)$ where $\Phi^a (z,y) \equiv x^a(z,y)/\kappa$
are adjoint Higgs fields of mass dimension defined by the covariant coordinates \eq{cov-coord}.

One can show, on the one hand, that, adopting the matrix representation \eq{op-matrix},
the $U(1)$ gauge theory on ${\bf R}^d_C \times {\bf R}^{2n}_{NC}$ is exactly mapped
to the $U(N  \to \infty)$ Yang-Mills theory on $d$-dimensional commutative space ${\bf R}^d_C$:
\begin{eqnarray} \label{matrix-action}
S_B &=& -\frac{1}{4 g^2_{YM}} \int d^D X
(F_{MN} - B_{MN}) \star (F^{MN} - B^{MN}) \nonumber \\
&=& - \frac{(2\pi\kappa)^{\frac{4-d}{2}}}{2\pi g_s} \int d^d z
{\rm Tr} \left(\frac{1}{4} F_{\mu\nu} F^{\mu\nu}  +
\frac{1}{2} D_\mu \Phi^a D^\mu \Phi^a  \right. \nonumber \\
&& \hspace{3cm} \left. - \frac{1}{4} [\Phi^a, \Phi^b]^2 \right),
\end{eqnarray}
where $F_{MN} (X) = \partial_M A_N - \partial_N A_M -i [ A_M, A_N]_\star \in U(1)$
and $D_\mu = \partial_\mu - i A_\mu (z) \in U(N  \to \infty)$.
Note that the 10-dimensional NC $U(1)$ gauge theory on ${\bf R}^{4}_C \times {\bf R}^{6}_{NC}$
is equivalent to the bosonic part of 4-dimensional ${\cal N} =4$
supersymmetric $U(N)$ Yang-Mills theory. Therefore it should not be so surprising that
a $D$-dimensional gravity could be emergent from the $d$-dimensional
$U(N  \to \infty)$ gauge theory in Eq.\eq{matrix-action},
according to the large N duality or AdS/CFT correspondence \ct{ads-cft}.

On the other hand, according to the map \eq{vector-map}, the $D$-dimensional NC $U(1)$
gauge fields $A_M (z,y) =(A_\mu, \Phi^a) (z,y)$ can be regarded as gauge fields
on ${\bf R}^d_C$ taking values in the Lie algebra of volume-preserving vector fields on
a $2n$-dimensional manifold $X$, i.e., the gauge group $G = SDiff(X)$:
\be \la{local-vector}
A_\mu(z) = A_\mu^a(z,y) \frac{\partial}{\partial y^a}, \qquad
\Phi_a(z) = \Phi_a^b(z,y) \frac{\partial}{\partial y^b}.
\ee
It turns out \ct{ward} that $f^{-1}(D_1, \cdots, D_d, \Phi_1, \cdots, \Phi_{2n})$
forms an orthonormal frame and hence defines a metric
on ${\bf R}^d_C \times X$ with a volume form $\nu = d^d z \wedge \omega$
where $f$ is a scalar, a conformal factor, defined by
$f^2 = \omega (\Phi_1, \cdots, \Phi_{2n})$ (see also \ct{cho}):
\begin{equation} \label{Ward}
ds^2 = f^2 \eta_{\mu\nu} dz^\mu dz^\nu + f^2 \delta_{ab}
V^a_c V^b_d (dy^c - {\bf A}^c) (dy^d - {\bf A}^d)
\end{equation}
where ${\bf A}^a = A^a_\mu dz^\mu$ and $V^a_c \Phi_b^c = \delta^a_b$.

The emergent gravity from NC field theories reveals a radically different picture
from Einstein gravity in the sense that gravity is not a fundamental force but a collective
phenomenon emerging from NC (or non-Abelian) gauge fields. (Although we are here confined
to NC $U(1)$ gauge theories, it was recently suggested \ct{harold} that
a NC $U(n)$ gauge theory should be interpreted as an $SU(n)$ gauge theory coupled to gravity.)
So it is inviting to ponder on the CC problem from the picture of the emergent gravity.
In order to address the problem with a new light, we properly change our question from
`why is the vacuum energy (almost) zero ?' to `why is the vacuum not gravitating ?', as
the tiny observed value \eq{cc-obs} already drops a hint.

A remarkable picture in emergent gravity is that spacetime is also emergent from gauge
field interactions \ct{est-seiberg}. Note that the metric
\eq{Ward} becomes flat when all fluctuations are turned off, say,
$(D_\mu, \Phi^a) = (\partial_\mu, y^a/\kappa)$. In other words, the
flat spacetime as a vacuum geometry is emergent from the uniform
condensation of gauge fields, i.e.,
\be \la{vacuum-condensation}
\langle B_{ab} \rangle_{\rm{vac}}= (\theta^{-1})_{ab},
\ee
which defines the NC C*-algebra \eq{nc-spacetime} \ct{t-duality}.

Therefore a flat spacetime is indeed originated from the uniform
condensation of energy in a vacuum.
A more crucial point is that the action \eq{matrix-action} is
invariant under the shift transformation by a constant like
Eq.\eq{shift-symm}. This shift effectively changes
the background \eq{vacuum-condensation} from $B$ to $B^\prime$ or
from $\theta$ to $\theta^\prime$. However NC gauge theories
for $\theta$ and $\theta^\prime$ are physically equivalent, i.e.,
$S_B \cong S_{B^\prime}$, which is precisely the Seiberg-Witten equivalence
between NC field theories \ct{sw}. Furthermore the vacuum geometry emerging from both
$\theta$ and $\theta^\prime$ is equally a flat spacetime
as long as they are constant, as Eq.\eq{vector-map} clearly shows.
Hence the vacuum energy such as Eq.\eq{vacuum-energy} will only appear as readjusting
the vacuum \eq{vacuum-condensation} without affecting any physical results.
In other words, any kinds of constant vacuum energy, previously identified with the CC,
are universally gauged away; they are used to make a flat spacetime.
So we arrive at a critical point for the CC problem that {\it the vacuum energy is not
gravitating} unlike as Einstein gravity. The same conclusion was already achieved
and was deemed to be critical for the resolution of the CC problem
in a perceptive work \ct{pady2}.

The emergent gravity reveals an intriguing picture about the origin of flat spacetime.
A flat spacetime is not free gratis, but a result of the condensation of
the Planck energy \eq{vacuum-energy}, the maximum energy in Nature, in a vacuum.
This novel dynamical origin of the vacuum \eq{vacuum-condensation} may explain the reason
why the flat spacetime as well as the Lorentz symmetry
as its spacetime symmetry are so robust against any perturbations.

Moreover, the vacuum \eq{vacuum-condensation} triggered by the Planck energy condensations
describes the NC spacetime whose defining algebra \eq{nc-spacetime} is equivalent to that of
harmonic oscillators as illustrated in Eq.\eq{matrix-basis}.
Thus the spacetime corresponds to a vast accumulation of harmonic oscillators
and so behaves as a fluid with negative pressure.

Now the problem is how to explain the small nonzero value \eq{cc-obs}
for the observed CC. A natural guess is to consider vacuum
fluctuations $\delta B_{ab}$ around the primary background \eq{vacuum-condensation}.
First we notice that NC spacetime leads to a perplexing mixing
between short (UV) and large (IR) distance scales \ct{uv-ir}. Thereby a UV fluctuation in
the NC spacetime \eq{nc-spacetime} whose natural scale is, as we know, the Planck scale
$L_P$ is necessarily paired with a corresponding IR scale $L_H$. A
simple dimensional analysis shows
that $|\delta B_{ab}| \sim 1/ L_P L_H$. Thus we estimate the
energy density of the vacuum fluctuation is of the order
\be \la{vacuum-fluctuation}
\rho_{\rm{vac}} \sim |\delta B_{ab}|^2 \sim \frac{1}{L^2_P L^2_H}.
\ee
Note that these vacuum fluctuations are not uniform but of size
$L_H$. It is natural to identify the IR scale $L_H$ with the size of
cosmic horizon in our Universe. Then Eq.\eq{vacuum-fluctuation}
coincides with the dark energy \eq{cc} \ct{pady2}.
Furthermore, numerically, with $L_P \sim 10^{-18} GeV^{-1}$ and
$L_H \sim 10^{42} GeV^{-1}$, one obtains $\rho_{\rm{vac}} \sim
(10^{-12} GeV)^{4}$ to be in agreement with the
observed value \eq{cc-obs}. This agreement up to a
factor is good enough since it would be pretentious to simulate an
exact factor at this stage.

The emergent gravity here would be the first scenario for the CC
problem showing microscopically how to gauge away huge zero-point
energies such as Eq.\eq{vacuum-energy}. Indeed we notice that the emergent gravity
remarkably realizes all criteria in \ct{pady1,pady2,volovik} suggested
as a possible solution for the CC problem. Nevertheless there are
still several important open issues to be clarified in the future.

Our picture for the CC problem, particularly, seems to imply that explosive
inflation era lasted roughly $10^{-33}$ seconds at the beginning of
our Universe corresponds to a dynamical process for an instantaneous
condensation of vacuum energy \eq{vacuum-energy} or \eq{vacuum-condensation}
to enormously spread out a flat spacetime. However it is not clear
how to describe this process
in terms of the NC (or matrix) action \eq{matrix-action}.

{\bf Acknowledgements} \\
I would like to thank Harald Dorn, Qing-Guo Huang and Bumseok Kyae
for valuable discussions. I am also grateful to Jan Plefka for his
generous support during the period of my extended stay at the QFT
group in Humboldt Universit\"at zu Berlin.


\begin{thebibliography}{55}
\bibitem{t-duality} More precisely, $\partial_\mu$ in $D_\mu$ must also be regarded as a
background. Indeed $\partial_\mu$ and $y^a/\kappa$ are related to
each other by a `matrix T-duality' \ct{hsy6}; $i D_\mu
\leftrightarrow \Phi^a$.
\end{thebibliography}
\end{document}